\documentclass{article}
\usepackage{emulateapj,apjfonts}
\input psfig

\newcommand{\etal}{{et~al.}}

\newcommand{\ts}{\thinspace}

\newcommand{\sd}{\newdimen\sa \sa=.1em  \ifmmode $\rlap{.}$''$\kern -\sa$
                                \else \rlap{.}$''$\kern -\sa\fi}

\begin{document}

\lefthead{NGC~4486A}
\righthead{Kormendy~\etal}

\title{THE NUCLEAR DISK IN THE DWARF ELLIPTICAL GALAXY 
NGC~4486A\altaffilmark{1}}

\author{John Kormendy\altaffilmark{2,3}, Karl Gebhardt\altaffilmark{2,3}, 
F.~Duccio Macchetto\altaffilmark{4}, W.~B.~Sparks\altaffilmark{4}}

\altaffiltext{1}{Based on observations with the NASA/ESA {\it Hubble
Space Telescope}, obtained at the Space Telescope Science Institute,
which is operated by the Association of Universities for Research in
Astronomy, Inc.~(AURA), under NASA contract NAS5-26555.}

\altaffiltext{2}{Visiting Astronomer, Canada--France--Hawaii Telescope, 
operated by the National Research Council of Canada, the Centre
National de la Recherche Scientifique of France, and the University of
Hawaii.}

\altaffiltext{3}{Department of Astronomy, University of Texas, Austin,
Texas 78712; kormendy@astro.as.utexas.edu, gebhardt@astro.as.utexas.edu}
 
\altaffiltext{4}{Space Telescope Science Institute, 3700 San Martin 
Drive, Baltimore, MD 21218}

\begin{abstract}

Many ellipticals contain nuclear disks of dust and gas.  Some
ellipticals contain nuclear disks of stars that are distinct from the
rest of the galaxy.  We show that the dwarf E2 galaxy NGC 4486A
contains both -- it is a ``Rosetta stone" object that tells us how
nuclear disks evolve.  Its properties suggest that, as accreted gas
dissipates and settles toward the center, it forms stars and builds a
stellar disk.  Secular growth may explain not only the most distinct
nuclear disks such as those in NGC 4486A but also some of the disky
distortions that are commonly seen in elliptical galaxies.  That is,
density distributions may grow secularly cuspier.  This would result
in chaotic mixing of stellar orbits in phase space and would tend to
make an elliptical galaxy evolve toward a more nearly axisymmetric
shape.

\end{abstract}


\section{Introduction}

The fraction of all elliptical galaxies that are known to contain dust
has risen dramatically as the resolution of our observations has
improved and as digital detectors have made it possible to see subtle
absorption features superposed on steep brightness gradients.
Kormendy \& Djorgovski (1989) review ground-based observations that
reveal dust, often in well defined disks, in at least
20\ts--\ts40\ts\% of all ellipticals (Hawarden
\etal 1981; Sadler \& Gerhard 1985a, b; Sparks \etal 1985; Ebneter \&
Balick 1985; Djorgovski \& Ebneter 1986; Kormendy \& Stauffer 1987;
Ebneter \etal 1988).  More recent ground-based surveys find dust in
even higher fractions of early-type galaxies (e.{\ts}g.,Ferrari \etal
1999).  The {\it Hubble Space Telescope\/} ({\it HST\/}) reveals dust
disks in spectacular detail.  It has been used to discover additional
dust disks that are too small to be seen from the ground (see Jaffe
\etal 1994; van Dokkum \& Franx 1995, Ford \etal 1998, Jaffe \etal
1999, Tomita \etal 2000, Tran \etal 2001 for reviews).  Also, even
prototypical bulges like M{\ts}31 are frequently riddled with dust
(Johnson \& Hanna 1972; Kent 1983; McElroy 1983) and ionized gas
(Ciardullo \etal 1988).  Kinematic axes that are misaligned with
respect to photometric axes show that many of these disks are accreted
(see Kormendy \& Djorgovski 1989 for a review).

Nuclear disks of stars are also seen in some ellipticals and
bulges.~Examples in bulges are found in NGC 3115 (Kormendy \etal
1996a) and NGC 4594 (Kormendy {\it et al.} 1996b).  Ellipticals with
stellar nuclear disks include NGC 4570 (van den Bosch \etal 1994) and,
most spectacularly, NGC 3706 (Lauer \etal 2001).  Clues to the
formation of these disks are the main subject of this paper.

The E3 galaxy NGC 5845 contains both a dust disk and an associated,
nuclear stellar disk.  Kormendy \etal (1994) emphasize that it
provides clues to how dust disks evolve. They suggest that, as
accreted gas dissipates and settles toward the center, it forms stars
and builds a stellar disk.  Secular growth may explain not only the
most distinct nuclear disks such as those in the above galaxies but
also some of the disky distortions that are commonly seen in
elliptical galaxies (e.{\ts}g., Bender \etal 1989).  That is, density
distributions may grow secularly cuspier.  This is important because
it means that the distribution of orbits and hence the degree of
triaxiality can evolve (see Merritt 1999 for a review).

NGC~4486A now provides a second example of associated dust and stellar
disks.  At a distance of 16 Mpc, it is an absolute magnitude $M_B
\simeq -17.6$ dwarf elliptical companion of M{\ts}87 (NGC~4486).
Binggeli, Sandage, \& Tammann (1985) classify it as E2.  The galaxy is
rarely studied because there is a bright star only 2\sd5 from its
center.  This star is ideal for ground-based adaptive optics (AO)
correction.  JK and KG obtained $K$-band AO images using the
Canada-France-Hawaii Telescope (CFHT) as part of a general
investigation of Virgo Cluster ellipticals and found that the galaxy
contains an edge-on nuclear disk.  In the infrared, there is no
obvious dust, but the center is offset from the outer isophotes by
$\sim 0\sd05$ along the minor axis.  This suggested that there may be
a dust lane.  Fortunately, FDM and WBS had obtained {\it HST\/} $V$-
and $I$-band WFPC2 images as part of an unrelated program.  These
images show an edge-on dust disk.  The following sections discuss the
properties and implications of the nuclear disks of NGC 4486A.

\section{ CFHT ADAPTIVE OPTICS IMAGE}

Figure 1 shows a $K$-band AO image obtained with the CFHT Adaptive
Optics Bonette (PUEO) and KIR camera (Rigaut \etal 1998).  The image
scale is 0\sd035 pixel$^{-1}$.  The images were obtained on 1999 March
6 in photometric conditions.  The total exposure time was 24 min,
taken in three groups of 24, 20 s exposures interleaved with sky
exposures.  The uncorrected seeing was $\sim 0\sd7$ FWHM.  Given the
very bright guide star only 2\sd5 from the galaxy, the AO compensation
is excellent; the stellar FWHM is $\sim 0\sd12$ and the Strehl ratio
is 0.6.  Two diffraction rings are visible around the star in Figure
1.  A galaxy-subtracted image of this star was used as a point-spread
function (PSF) for Lucy-Richardson deconvolution; 40 iterations
eliminated most of the diffraction pattern and reduced the FWHM to
0\sd07.  Further details of the observations are given in Gebhardt \&
Kormendy (2001).

Figure 1 shows a surprising result: this elliptical galaxy contains a
prominent, almost edge-on nuclear disk of stars.  This does not mean
that the galaxy is an S0.  S0 disks are at large radii, outside most
of the bulge (Sandage, Freeman, \& Stokes 1970).  There is no evidence
for an outer disk here.  In contrast, the disk in NGC 4486A is similar
to the nuclear disks that are found interior to and not connected with
the outer disks in the S0 galaxy NGC 3115, the Sa galaxy NGC 4594, and
some giant ellipticals~(\S\ts1).

The outer part of the disk is symmetric, but the brightest pixels near
the center are displaced by $\sim 0\sd05$ along the minor axis.  This
suggested that the galaxy may contain a dust lane that is almost
negligible in the infrared.

\section{{\it HUBBLE SPACE TELESCOPE} WFPC2 IMAGES}

The presence of dust is clearly established \hbox{by Figure 2,} which
shows WFPC2 images obtained in the $V$ and $I$ bandpasses (F555W and
F814W).  These images were processed using the standard STScI
calibration pipeline.  Each is the sum of two, 500 s exposures that
were combined using the iraf/stsdas task ``crrej'' to eliminate cosmic
rays.

These images were obtained for an unrelated program, but NGC 4486A is
fortuitously positioned close to the center of the PC.  Figure 2 shows
that the dust lane is very well defined and has a sharp outer edge.
It is embedded in the inner part of the stellar disk, but the stellar
disk extends to twice the radius of the dust disk.

\section{ORIGIN OF NUCLEAR DISKS OF STARS}

The morphology of the nuclear disk is best illustrated by constructing
a three-color image from the $V$-, \hbox{$I$-,} and $K$-band exposures
(Figure 3).  NGC~4486A is important because it contains both a stellar
and a dust disk.  They are regular and clearly associated.  As in NGC
5845, this provides a key clue toward our understanding of how nuclear
disks evolve. It does not tell us how the dust originates.  Other
evidence demonstrates that dust is commonly accreted (Kormendy \&
Djorgovski 1989), but it does not exclude that it sometimes has an
internal origin.  For convenience, we refer to the dust as accreted
material. Whatever its origin, it migrates quickly toward the center.
There, gas and dust settle into a disk that, in equilibrium, is
oriented perpendicular to the short axis if the object is a spheroid
or perpendicular to the shortest or longest axis it the object is
triaxial (Heiligman \& Schwarzschild 1979).  Objects like NGC 4486A
and NGC 5845 illuminate what happens next.  Gas likes to make stars
when it gets crunched, and funneling the gas into the compact center
is a natural way to crunch it.~With Kormendy \etal (1994), we suggest
that the gas and dust turn into stars and build a stellar disk.

The possible caveat is this: if the nuclear disk of stars predates the
accretion event because it formed with the galaxy, then it helps to
define a gravitational potential that will make newly acquired gas
settle into the same disk plane and look like it is associated with
\hbox{the stellar disk.}  On the other hand, if the stellar disk
formed recently from the accreted material, then it may be observably
younger than the rest of the galaxy.  The best way to test this is
spectroscopy.  This is in progress. Here, we can look for a color
difference between the nuclear disk and the bulge.

Figure~4 shows a $V - I$ color image of NGC 4486A.  The blue PSF halo
produced by scattered light clobbers the southern part of the stellar
disk. But Figure 4 shows (1) that light from the dust lane is very
red, and (iii) that the stellar disk outside the dust lane is
marginally bluer than the rest of the galaxy.  After a convolution of
the $V$ image to match it to the spatial resolution of the $I$ image,
the difference in color is $\Delta (V - I) = 0.06$ mag.  Broad-band
colors are not a unique indicator of age, but this result supports the
hypothesis that gas disks build stellar disks near the centers of
bulges and elliptical galaxies.

\section{SECULAR GROWTH OF CUSPY DENSITY DISTRIBUTIONS IN BULGES AND 
ELLIPTICAL GALAXIES}

The implication of these results is that the density distributions of
bulges and elliptical galaxies grow cuspier even after the events that
formed them.  This has practical consequences; it makes the search for
supermassive black holes easier (because the brightness is higher) and
more reliable (because the dynamics are dominated by rotation, so
velocity anisotropy is less important).  It also adds noise to the
correlations between central and global properties of elliptical
galaxies.  These suggest that there are two kinds of ellipticals, (1)
slowly rotating, anisotropic, triaxial, boxy-distorted ellipticals
with cuspy cores and (2) rapidly rotating, nearly isotropic and
spheroidal, disky-distorted, coreless ellipticals (Bender 1988; Bender
\etal 1989; Nieto, Bender, \& Surma 1991; Kormendy \etal 1994;
Kormendy \& Bender 1996; Faber \etal 1997; Tremblay \& Merritt 1996).
In particular, it may account for exceptions to the dichotomy, like
NGC 1316 = Fornax A (Schweizer 1980; Kormendy 1987; Faber \etal 1997)
and NGC 4621 (Kormendy \etal 1994).  Both galaxies are unusually cuspy
for their high luminosities.

However, the main reason why the growth of cuspiness is important is
that it results in secular evolution of the overall structure of
elliptical galaxies.  Merritt and collaborators (see Merritt 1999 for
a review) have shown that the growth of cuspy density distributions
makes the orbit structure evolve even at radii substantially outside
the radius out to which the cusp dominates the potential.  Stars on
box orbits pass close to the center and scatter off of the density
cusp.  This results in chaotic mixing, which redistributes the orbits
in phase space and makes the galaxy evolve rapidly toward axisymmetry.
One driving agent is the growth of nuclear black holes.  Increasing
the cuspiness of the stellar mass distribution is another.

What the present observations do not tell us is how much growth in
density is typical.  The frequent observation of dust disks suggests
that some evolution happens to nearly every elliptical.  But
timescales and mass deposition rates are unknown.  We need to know how
much gas and dust is observed and to investigate the ages of nuclear
disk stars to constrain the timescale of disk growth.

What the present results provide is a proof of concept.  They imply
that evolution does take place, and that the fate of the gas and dust
disks that we see in so many elliptical galaxies is to turn into
nuclear disks of stars.


\begin{figure*}[t]
\centerline{\psfig{file=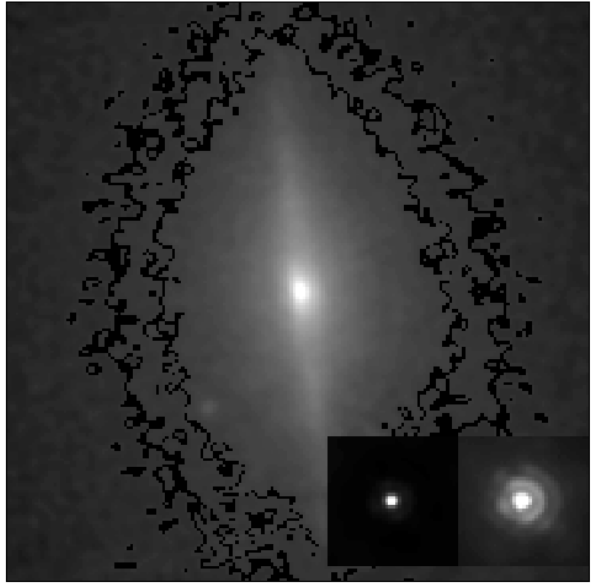,width=10cm,angle=0}}
\vskip -80pt
\figcaption[kormendy.fig1.ps]{NGC~4486A: $K$-band adaptive optics 
image obtained with the CFHT after 40 iterations of Lucy
deconvolution.  The field size is 6\sd4; north is up and east is at
left.  Brightness is proportional to the square root of intensity, and
two isophotes are blacked out.  For display purposes, a region around
the bright star has been divided by 100 to approximately match the
peak brightness in the galaxy.  The FWHM is 0\sd07.  To its right, an
inset shows the star before deconvolution.
\label{fig1}}
\end{figure*}



\begin{figure*}[b]
\centerline{\psfig{file=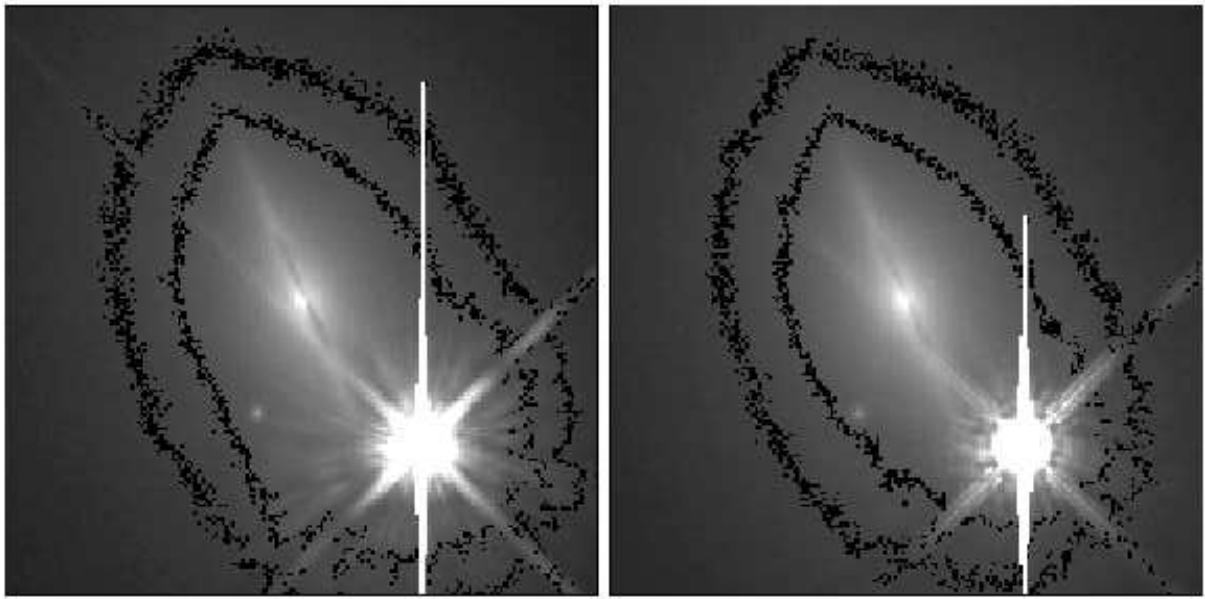,width=22.0cm,angle=-90}}
\vskip -150pt
\figcaption[kormendy.fig2.ps]{NGC~4486A: WFPC2, $V$-band 
({\it left\/}) and $I$-band ({\it right\/}) images.  The field size is
$8^{\prime\prime}$.  Brightness is proportional to the square root of
intensity, and two isophotes are blacked out.
\label{fig2}}
\end{figure*}


\clearpage


\begin{figure*}[t]
\centerline{\psfig{file=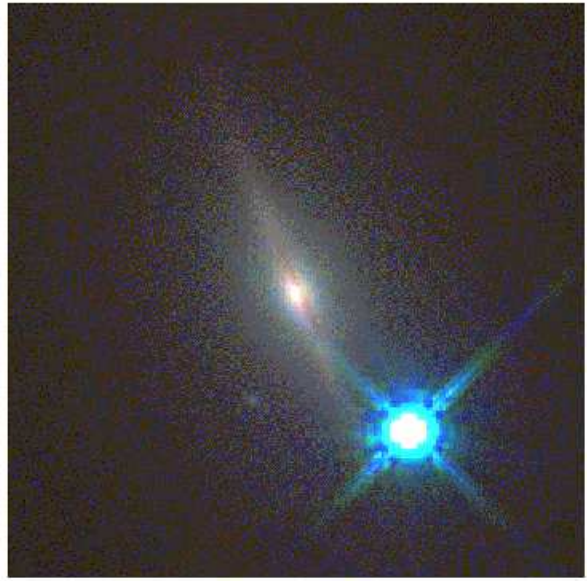,width=13.0cm,angle=0}}
\vskip -120pt \figcaption[kormendy.fig3.ps]{Three-color image of
NGC~4486A constructed by coding the $V$ image as blue, the $I$ image
as green, and the $K$ image as red.  The field size is
$8^{\prime\prime}$.  The intensity stretch is linear.
\label{fig3}}
\end{figure*}



\begin{figure*}[b]
\centerline{\psfig{file=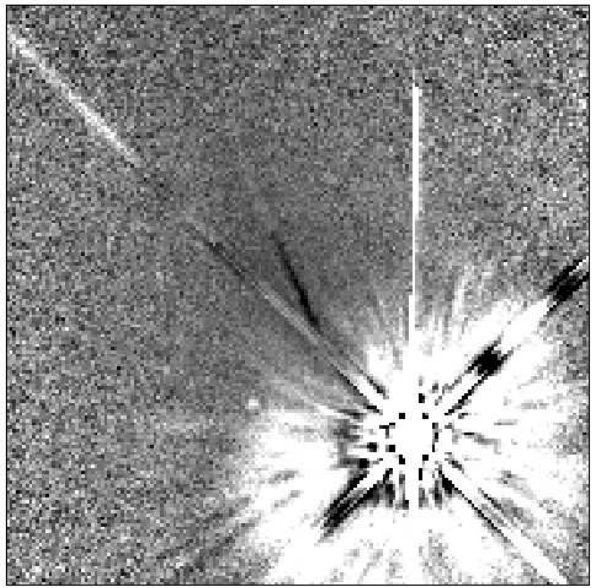,width=10.0cm,angle=0}}
\vskip -90pt \figcaption[kormendy.fig4.ps]{$V - I$ color image of
NGC~4486A.  Black corresponds to $V - I = 1.7$ and white corresponds
to $V - I = 1.2$.  The bulge color is $V - I \simeq 1.41$ The stellar
disk is $\Delta (V - I) = 0.06$ bluer than the bulge.  The field size
is $8^{\prime\prime}$.
\label{fig4}}
\end{figure*}


\end{document}